\documentclass[11pt]{article}
\usepackage{array}
\usepackage{graphicx}
\usepackage{cite}
\usepackage{textpos}
\usepackage{bbold}
\usepackage{adjustbox}

\usepackage[normalem]{ulem}
\usepackage{amsmath}
\newcommand{\stkout}[1]{\ifmmode\text{\sout{\ensuremath{#1}}}\else\sout{#1}\fi} 

\title{Probability of Presence Versus $\psi^*(x, t) \psi (x, t)$ }
\author {Frank Wilczek  \\
\small\it Center for Theoretical Physics, MIT, Cambridge, MA 02139 USA; \\
\small\it T. D. Lee Institute and Wilczek Quantum Center, \\
\small\it Shanghai Jiao Tong University, Shanghai, China;\\
\small\it Arizona State University, Tempe, AZ, USA; \\
\small\it Stockholm University, Stockholm, Sweden  \\
{} \\
Zara Yu \\
\small\it Department of Physics, MIT, Cambridge, MA 02139 USA}

\begin{document}

\maketitle

\begin{textblock*}{5cm}(10cm,-10.2cm)
\fbox{\footnotesize MIT-CTP/5719}
\end{textblock*}

\begin{abstract}
Postulating the identification of $\psi^*(x, t) \psi(x,t)$ with a physical probability density is unsatisfactory conceptually and overly limited practically.  For electrons, there is a simple, calculable relativistic correction proportional to $\nabla \psi^* \cdot \nabla \psi$.  In particular, zeroes of the wave function do not indicate vanishing probability density of presence.  We derive a correction of this kind from a Lagrangian, in a form suitable for wide generalization and use in effective field theories.  Thus we define a large new class of candidate models for (quasi-)particles and fields, featuring modified {\it kinetic\/} terms.  We solve for the stationary states and energy spectrum in some representative problems, finding striking effects including the emergence of negative effective mass at high energy and of localization by energy.
\end{abstract}

\medskip

\bigskip


Textbooks on quantum mechanics, when they come to making contact with the empirical world, tend to postulate that there is such a thing as a particle that has amplitudes to be  at different times and places -- i.e., a wave-function 
$\psi(x,t)$ -- and that $\psi^*\psi$ represents the (unnormalized) probability for it to ``be there then''.  (See for example these three excellent modern texts: \cite{sakurai, griffiths, shankar}.) Here we will examine that probability postulate critically.  We will argue that the choice  $\psi^*\psi$ can be modified in ways that are fully consistent with the basic principles of quantum theory.  They entail  associated changes in the form of the Schr\"odinger equation and of conservation laws, and in the energy spectrum.  

We will derive all these features within the framework of Lagrangian dynamics, based on addition of formally simple (local, low-dimension, highly symmetric) terms to the standard kinetic energy term.   Such terms are good candidates for inclusion in effective theories of (quasi-)particles.  Plausibly, they will occur with significantly large coefficients in appropriate materials.   Striking qualitative and quantitative consequences can ensue, as we demonstrate in simple representative problems.  

\section{Conceptual Critique}

\subsection{Structure of Points}\label{point_structure}

At a formal level, the issue raised here arises in the following way.  If we assume that there is a dynamical variable $x$ corresponding to the position of a particle, then there will be states $| x \rangle $ that diagonalize it, and for a general state the expansion $| \psi \rangle = \int dx \, \psi (x) |x \rangle$.   This general framework does not yet supply enough structure, however, for us to calculate the probability density for finding the particle at $x_0$, {\it viz}. 
$ \langle \psi |  \delta (x-x_0)  | \psi \rangle  $.  To do that, according to 
\begin{equation}\label{probability_expression}
\langle \psi |  \delta (x-x_0)  | \psi \rangle  ~=~ \int \int dx_2 dx_1 \psi^*(x_2) \delta (x_1 -x_0) \psi(x_1)  \langle x_2 | x_1 \rangle \, , 
\end{equation}
we need to have $\langle x_2 | x_1 \rangle$.  The standard prescription 
\begin{equation}\label{standard}
\langle x_2 | x_1 \rangle_s ~=~ \delta (x_2 - x_1) 
\end{equation}
leads to the standard consequence $\langle \psi |  \delta (x-x_0)  | \psi \rangle_s= \psi^* (x_0) \psi(x_0)$.    
In this language, looking ahead, Eqn.\,(\ref{corrected_density}) corresponds to the choice
\begin{equation}\label{inner_product}
\langle x_2 | x_1 \rangle  ~=~ \biggl(1 - \frac{\nabla^2_{x_2}}{4m^2}\biggr)\delta (x_2 - x_1) \, . 
\end{equation}
Comparing the form Eqn.\,(\ref{inner_product}) with Eqn.\,(\ref{standard}), we may say that points, as perceived by an electron, have acquired interior structure.  The unconventional normalization Eqn.\,(\ref{inner_product}) is connected, through the completeness relation
\begin{equation}
\int [dx] \langle x_2 | x \rangle \langle x| x_1 \rangle  ~=~ \langle x_2 | x_1 \rangle \, , 
\end{equation}
to the unconventional measure 
\begin{equation}
[dx] ~=~ dx \, \biggl(1 - \frac{\nabla_{x}^2}{4m^2}\biggr)^{-1} \, .
\end{equation}

Here we should remark that in Eqn.\,(\ref{probability_expression}) we have assumed that the state conjugate, in the sense of Hilbert space, to the state described by the wave function $\psi(x)$ is described by $\psi^*(x)$.  The physically meaningful content of this association is to say that the dual of $\psi(x) | x \rangle$ is $\psi^*(x)\langle x|$.   Thus, we can trade our unconventional normalization of $\langle x |$ for an unconventional implementation of conjugate wave-functions.

\subsection{Measurement}

In other parts of the textbooks, it is often emphasized that we should be careful about assigning physical reality to things that we don’t measure.  In that spirit: What sort of measurement corresponds to determining the probability that an electron will be found at a given space-time point?  

It is difficult for an answer to be better defined than the question it responds to.  But here there is a useful answer that leads us back to the preceding choices.  That is, we recognize that many practical ways of ``locating an electron'' involve sensing  its interaction with electromagnetic fields.  From this perspective, the position of an electron is a theoretical construct for describing the electromagnetic response of systems that can be usefully modeled based on a theory built up from electron particle variables.  In this context, it is natural to identify electron density with the density of electric charge (more accurately, the part of the electric charge operator that we ascribe, in the model, to electrons).   The electric charge density operator is uniquely determined, so it does provide a definite answer to our question.  Other answers might be appropriate to other ways of implementing the question experimentally but if we're looking for a specific answer then charge density appears to be the most salient possibility.  (We should not expect the answer to a question to be better defined than the question.)  We will refer to this answer as ``probability of presence'', as opposed to simply ``probability density'', to emphasize how it is to be interpreted physically.

\subsection{Philosophy of Quantum Modeling}

The construction of quantum-mechanical models of physical phenomena can proceed at different levels.  At present, it appears that the so-called Standard Model provides, in principle, an adequate foundation  for most applications of physics to the natural world, outside of early universe cosmology.  The Standard Model, based on relativistic quantum field theory and local gauge symmetry, embodies the general principles of quantum mechanics and the special theory of relativity, makes quantitative predictions that show impressive agreement with experimental measurements on fundamental processes and has good ultraviolet properties (i.e., allows consistent extrapolation to high energy). It retains those excellent features, other than good ultraviolet behavior, if we incorporate gravity as described by general relativity, using minimal couplings.  String theory promises to provide a framework through which to incorporate gravity while keeping good ultraviolet behavior, but as yet it does not supply algorithms for quantitative comparison with nature.  

But these ``fundamental'' theories, when applied to any but the simplest physical processes, provide an unwieldy description that is impractical to use.   For practical purposes, we build models that incorporate fewer basic principles and allow more flexible choices for dynamical variables.  A philosophy that has proved fruitful in many modern applications is to explore systematically among consistent implementations the general principles of quantum theory together with appropriate symmetries, locality, and some criterion of simplicity.   In this way we hope to provide plausible candidate models for the approximate description of more complex systems, such as molecules or materials.   It has also been fruitful to move in the opposite direction, sculpting material systems ({\it metamaterials\/}, broadly defined)  that realize interesting models.

Above we have spoken of electrons, and invoked fundamental electrodynamics in empty space.  Specifically, Eqn.\,(\ref{inner_product}) was reverse-engineered to reproduce the result of approximating the Dirac theory, regarded as fundamental.  (See Section \ref{dirac_correction}.) Intuitively, we might associate the ``spread'' of probability of presence associated with electrons in vacuum with their irreducible uncertainty in position, associated with their small-scale motion, or {\it zitterbewegung}, that has been integrated out in the non-relativistic description. In Appendix \ref{Dirac_reduction} we discuss how our Lagrangian approach to the non-relativistic limit of the Dirac equation reproduces the results of more traditional approaches.  

Correction terms of a similar mathematical form can be expected to arise in the quantum description of particles with extended structure, such as nucleons (for which the Compton and geometric sizes are comparable) or, with significantly larger coefficients, atoms and molecules.   This idea can be implemented mathematically, and leads us to a potentially significant expansion of the repertoire of quantum models that embody basic symmetry principles, locality, and simplicity, as we shall show.

In purely mathematical terms, one can view Eqn.\,(\ref{inner_product}) as exemplifying a wide class of possibilities for building quantum mechanical models, wherein the minimal $L^2(R)$ norm used in constructing the Hilbert space of wave functions (for particles on a line) is replaced by a Sobolev-style norm
\begin{equation}
\langle \psi | \psi \rangle ~=~  \int \, dx \, \sum\limits_{j= 0}^{n} \, a_j \, |\partial^{\, j}_x \psi |^2.
\end{equation}
Note that this is bilinear and, for suitable ranges of $a_j$, positive.
But to get useful models we need to bring in more structure, as we will now describe.

\section{Correction from Dirac Theory}\label{dirac_correction}

The relativistic (Dirac) theory for electrons suggests a specific correction to the conventional position probability density $\psi^*\psi$.  In the Dirac theory there is an underlying 4-component spinor $\Psi$, and a density 
\begin{equation}
\rho = \Psi^\dagger \Psi.
\end{equation}
Since this density is positive-definite, and associated with a conserved 4-current $j^\mu = \bar \Psi \gamma^\mu \Psi$, it is natural to associate $\rho$ with the probability density for finding an electron, and there is no reasonable alternative candidate with those properties.    In the non-relativistic limit $\Psi$ takes the form 
\begin{equation}
\Psi \approx \left(\begin{array}{c}\psi \\ \frac{\sigma \cdot p} {2m} \, \psi \end{array}\right)
\end{equation}
where $\psi$ is a two-component spinor.  In terms of $\psi$, then, the probability density is
\begin{equation}\label{corrected_density}
\rho \approx \psi^* \psi + \frac{1}{4m^2} \nabla \psi^* \cdot \nabla \psi.
\end{equation}

Eqn.\,(\ref{corrected_density}) includes a correction to the conventionally assumed expression.  The correction is of order $\frac{p^2}{m^2} \sim \frac{v^2}{c^2}$, and we can expect it to be small within most practical applications of non-relativistic quantum mechanics.  Still, it has the qualitatively and conceptually significant effects.  It removes zeroes in the probability distribution, since one cannot expect both $\psi$ and $\nabla \psi$ to vanish at the same place.   More generally, we can anticipate its quantitative significance will emerge at places where the wave function is small while its gradient is large, for example at the edge of a high, steep barrier.  As we shall see in examples, it also has striking effects on the highly excited states in bound state problems, which bring in large gradients due to orthogonality constraints.  Finally, let us note that in Weyl semi-metals and related materials, where one encounters the Dirac equation with a substantially smaller ``speed of light'', all these effects can be more significant quantitatively.

\section{Lagrangian Realization}

Effective theories based on Lagrangians allow us to realize the general principles of quantum theory and embody appropriate symmetries, including the number symmetries that help to specify the census of ingredients in the models.  Furthermore, they lend themselves to quantization using path integrals.  Now let us realize our conceptual considerations, and the motivating example of relativistic electron theory, within that framework.

To realize our correction term in this framework is significant for other reasons.  For one thing, it allows us to draw out all its implications, including the form of the associated 3-current, the form of the associated energy-momentum tensor density, and the form of the associated contribution to the equations of motion.  For another, it brings us into the spirit of Landau-Ginzburg theories, where we identify relevant parameters for the description of material systems based on their appearance in effective Lagrangians \cite{girvin}.   

\subsection{Mathematical Structure}

The conventional probability expression is identical with the charge density expression that arises for 
\begin{equation}
L ^{(0)}~=~ \frac{i}{2} \psi^* \stackrel{\leftrightarrow}{\partial_t} \psi.
\end{equation}
One might anticipate that a contribution to the Lagrangian of the form
\begin{equation}\label{new_probability_term}
L^{(1)}  ~\propto~ i \nabla \psi^*  \stackrel{\leftrightarrow}{\partial_t} \nabla \psi
\end{equation}
leads to a contribution to the location (i.e. charge) density of the form we're looking for.  That is correct, as we will now demonstrate directly in a way that applies more broadly.

As is traditional in problems of this kind, we regard the Lagrangian formally as a function of the fields acted upon by derivatives.  Thus the $U(1)$ phase invariance of $L$ is expressed as
\begin{equation}\label{phase_invariance}
0 = \frac{\delta L}{\delta\psi} \psi + \frac{\delta L}{\delta \partial_t \psi}\partial_t \psi + \frac{\delta L}{\delta \nabla \psi} \nabla \psi + \frac{\delta L}{\delta \partial_t \nabla \psi} \partial_t \nabla \psi - (\psi \rightarrow \psi^* )
\end{equation}
(including only the terms we will be using) and the equations of motion are
\begin{equation}\label{e_o_m}
0 = \frac{\delta L}{\delta \psi} - \partial_t  \frac{\delta L}{\delta \partial_t \psi} - \nabla \frac{\delta L}{\delta \nabla \psi} + \partial_t \nabla \frac{\delta L}{\delta \partial_t \nabla \psi}
\end{equation}
together with a similar equation with $\psi \rightarrow \psi^*$.

Inserting Eqn.\,(\ref{e_o_m}) into the first half of the right hand side of Eqn.\,(\ref{phase_invariance}) yields six terms, as follows
\begin{equation}
0 = (\nabla \frac{\delta L}{\delta \nabla \psi}   +  \partial_t \frac{\delta L}{\delta \partial_t \psi} -  \partial_t\nabla \frac{\delta L}{\delta \partial_t \nabla}) \psi + \frac{\delta L}{\delta \nabla \psi} \nabla \psi + \frac{\delta L}{\delta \partial_t \psi}\partial_t \psi + \frac{\delta L}{\delta \partial_t \nabla \psi} \partial_t \nabla \psi.
\end{equation}
The first and fourth terms combine as $\nabla (\frac{\delta L}{\delta \nabla\psi} \psi)$, while the second and fifth combine as $\partial_t ( \frac{\delta L}{\delta \partial_t \psi} \psi)$.  Finally, for the third and sixth we have
\begin{equation}
- \partial_t\nabla \frac{\delta L}{\delta \partial_t \nabla} \psi + \frac{\delta L}{\delta \partial_t \nabla \psi} \partial_t \nabla \psi = \partial_t (\frac{\delta L}{\delta \partial_t \nabla \psi} \nabla \psi)  - \nabla (\partial_t \frac{\delta L}{\delta \partial_t \nabla \psi} \psi).
\end{equation} 
Thus we express the first half of the right hand side of Eqn.\,(\ref{phase_invariance}) as 
\begin{equation}
\partial_t (\frac{\delta L}{\delta \partial_t \psi} \psi + \frac{\delta L}{\delta \partial_t \nabla \psi} \nabla \psi) + \nabla (\frac{\delta L}{\delta \nabla\psi} \psi - \partial_t \frac{\delta L}{\delta \partial_t \nabla \psi} \psi)
\end{equation}
and Eqn.\,(\ref{phase_invariance}) itself as
\begin{equation}\label{current_conservation}
0 = \partial_t ( \frac{\delta L}{\delta \partial_t \psi} \psi + \frac{\delta L}{\delta \partial_t \nabla \psi} \nabla \psi) + \nabla (\frac{\delta L}{\delta \nabla\psi} \psi - \partial_t \frac{\delta L}{\delta \partial_t \nabla \psi} \psi) - (\psi \rightarrow \psi^*).
\end{equation}

$L = L^{(0)}$ gives a presence density of the conventional form, while $L = L^{(1)}$ gives a presence density of the form $\nabla \psi^* \cdot \nabla \psi$.  In constructing quantum-mechanical models, of course, we can consider including both terms, together with other additions.  

\subsection{First Modified Lagrangian} 
Now let us work out general consequences of expanding the Schr\"odinger Lagrangian (i.e., the Lagrangian that has the Schr\"odinger equation as its equation of motion) to include the additional term discussed above, so that
\begin{equation}\label{modified_L}
L =  \frac{i}{2} \psi^* \stackrel{\leftrightarrow}{\partial_t} \psi + \frac{a}{2}  i \nabla \psi^*  \stackrel{\leftrightarrow}{\partial_t} \nabla \psi - V \psi^* \psi - \frac{1}{2m} \nabla \psi^* \cdot \nabla \psi.
\end{equation}
Here, to keep things appropriately simple, we have included a scalar field $V$.  It should not be interpreted as electric potential, since there is no gauge symmetry associated to it.  (See Section \ref{second_Lagrangian}.)  Also, we will allow $V$ to depend on space, but not on time. 

\subsubsection{Equation of Motion}
The modified Schr\"odinger equation -- i.e., the equation of motion derived from $L$ -- reads
\begin{align}\label{modified_Schr}
    0 &= \frac{\delta L}{\delta \psi^*} - \partial_t \frac{\delta L}{\delta \partial_t \psi^*} - \nabla \frac{\delta L}{\delta \nabla \psi^*}+ \partial_t \nabla \frac{\delta L}{\delta \partial_t \nabla \psi^*}\notag\\
    &= \biggl(-V \psi +  \frac{i}{2} \partial_t \psi\biggr) +  \partial_t \biggl(\frac{i}{2} \psi\biggr)\notag\\
  &- \nabla\biggl(-\frac{1}{2m} \nabla \psi + \frac{a}{2} \cdot i\nabla \partial_t \psi \biggr)-  \partial_t \nabla\biggl(\frac{a}{2} \cdot i\nabla \psi\biggr)\notag\\
    &= -V \psi  +  i\partial_t \psi  + \frac{1}{2m} \nabla^2 \psi  -  a \cdot i \partial_t \nabla^2 \psi.
\end{align}

\subsubsection{Conservation Laws}

\begin{enumerate}

\item Following the logic of the preceding section, we find the equation expressing local charge conservation
\begin{align}
    0  &= \partial_t(\psi^* \psi + a(\nabla \psi^* \cdot \nabla \psi))\notag\\
    &+ \nabla \biggl(\frac{i}{2m} (\nabla \psi^* \psi - \psi^*\nabla \psi) - a(\psi \partial_t \nabla \psi^* + \psi^* \partial_t \nabla \psi) \biggr). 
\end{align}
Thus, the quantity
\begin{equation}\label{presence_density}
\rho = \psi^* \psi + a(\nabla \psi^* \cdot \nabla \psi)
\end{equation}
has suitable properties to represent, after normalization, a probability distribution -- namely, it is positive definite and, given appropriate spatial boundary conditions, its integral over space is conserved.  On the other hand, the spatial integral  $$\int dx \psi^*(x, t) \psi (x, t)$$ is not independent of time, so $\psi^* \psi$ cannot be interpreted as a probability density.  This supports the use of $\rho$, rather than $\psi^* \psi$,  as the preferred measure of probability of presence.  

\item By adding the product of the equation of motion for $\psi$ with $\partial_t\psi^*$ to its complex conjugate, and re-organizing the terms, we obtain an equation that expresses local energy conservation:
\begin{eqnarray}
    0 &=&  \partial_t ( \psi^* V \psi + \frac{1}{2m} \nabla \psi^* \cdot \nabla \psi )  \\
    &+& \nabla \bigl( ia(\partial_t \psi^* \partial_t \nabla \psi - \partial_t \nabla \psi^* \partial_t \psi) - \frac{1}{2m} (\partial_t \psi^* \nabla \psi + \nabla \psi^* \partial_t \psi) \bigr). \nonumber
    \end{eqnarray}
The formal expression of the energy density 
\begin{equation}\label{energy_density}
    \varepsilon = \psi^* V \psi + \frac{1}{2m} \nabla \psi^* \cdot \nabla \psi
\end{equation}
is independent of $a$, but the formal expression for its flux 
\begin{equation}
    j_\varepsilon = ia(\partial_t \psi^* \partial_t \nabla \psi - \partial_t \nabla \psi^* \partial_t \psi) - \frac{1}{2m} (\partial_t \psi^* \nabla \psi + \nabla \psi^* \partial_t \psi)
\end{equation}
does depend on $a$.  Of course, in evaluating the energy density within a given state, we must take account of the normalization of the wave function, which can bring in $a$ dependence.

\item By adding the product of the equation of motion for $\psi$ with $\partial_k \psi^*$ to its complex conjugate, and re-organizing the terms, we obtain an equation that expresses the local change of momentum reacting to the force field $-\partial_kV$:
\begin{align}
     &-2\psi \psi^* \partial_k V  =  \notag \\
    &+ i\partial_t\biggl(\psi \partial_k \psi^* - \psi^* \partial_k \psi - a(\nabla \psi^* \partial_k \nabla \psi - \nabla \psi \partial_k \nabla \psi^*)\biggr)\notag\\
    &- \frac{1}{2m} \nabla (\psi^* \partial_k \nabla \psi - \nabla \psi \partial_k \psi^* + \psi \partial_k \nabla \psi^* - \nabla \psi^* \partial_k \psi)\notag\\
    &- ai\nabla(\partial_k \psi^* \partial_t \nabla \psi - \partial_k \psi \partial_t \nabla \psi^* + \psi \partial_k \partial_t \nabla \psi^* - \psi^* \partial_k \partial_t \nabla \psi).
\end{align}

From this, we identify the momentum density
\begin{equation}
    \pi_k = i\biggl(\psi \partial_k \psi^* - \psi^* \partial_k \psi - a (\nabla \psi^* \partial_k \nabla \psi - \nabla \psi \partial_k \nabla \psi^*)\biggr)
\end{equation}
and its flux 
\begin{align}
     j^{(\pi_k)}_l &= \frac{1}{2m} (\psi^* \partial_k \partial_l \psi - \partial_l \psi \partial_k \psi^* + \psi \partial_k \partial_l \psi^* - \partial_l \psi^* \partial_k \psi)\notag\\
    &+ ia (\partial_k \psi^* \partial_t \partial_l \psi - \partial_k \psi \partial_t \partial_l \psi^* + \psi \partial_k \partial_t \partial_l \psi^* - \psi^* \partial_k \partial_t \partial_l \psi).
\end{align}
The momentum flux density can be interpreted, following standard arguments, as a stress tensor
\begin{equation}
T_{kl} =   j^{(\pi_k)}_l.
\end{equation}

\end{enumerate}

\subsubsection{Examples: Effective Mass}

For stationary states, with $\psi(x, t) \propto e^{-iEt} \psi(x)$, the modified Schr\"odinger equation Eqn.\,(\ref{modified_Schr}) takes the form of the usual Schr\"odinger equation with a modified coefficient of the Laplacian term, i.e. an effective mass with
\begin{eqnarray}\label{eff_mass}
\frac{1}{2m_{\rm eff.}}~&=&~ \frac{1}{2m} - a E \nonumber \\
m_{\rm eff.} ~&=&~ \frac{m}{1-2amE}.
\end{eqnarray}
In cases where the energy levels of the ordinary Schr\"odinger equation have a simple analytical dependence on mass, this observation allows us to calculate the energy eigenvalues for the modified Schr\"odinger equation.  

Thus, for the harmonic oscillator with spring constant $k$ and mass $m$ we have the energy eigenvalues
\begin{equation}
\epsilon_n ~=~ (n+ \frac{1}{2})  \sqrt{\frac{k}{m}} 
\end{equation}
with $n\geq 0$, and we infer for the modified Schr\"odinger equation the energy eigenvalues
\begin{equation}
E_n ~=~ (n+ \frac{1}{2}) \sqrt{\frac{k(1-2amE_n)}{m}} 
\end{equation}
leading to 
\begin{equation}
E_n ~=~ \epsilon_n(\sqrt{1 + (\epsilon_n a m)^2} - \epsilon_n a m).
\end{equation}

\begin{figure}
  \centering
  \begin{tabular}{@{}c@{}}
    \includegraphics[width=10cm]{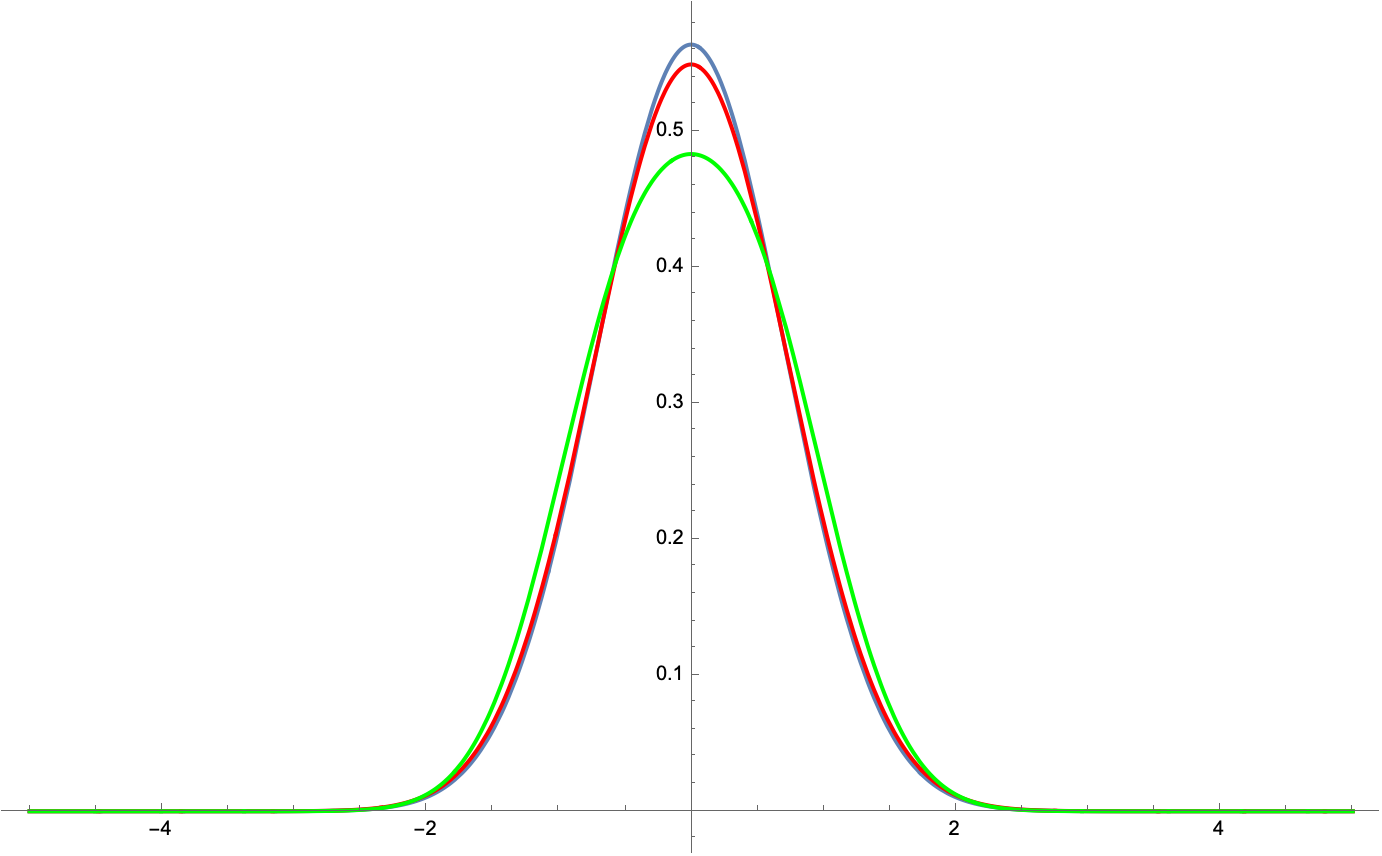} \\[\abovecaptionskip]
    \small (a) Ground state
  \end{tabular}

  \vspace{\floatsep}

  \begin{tabular}{@{}c@{}}
    \includegraphics[width=10cm]{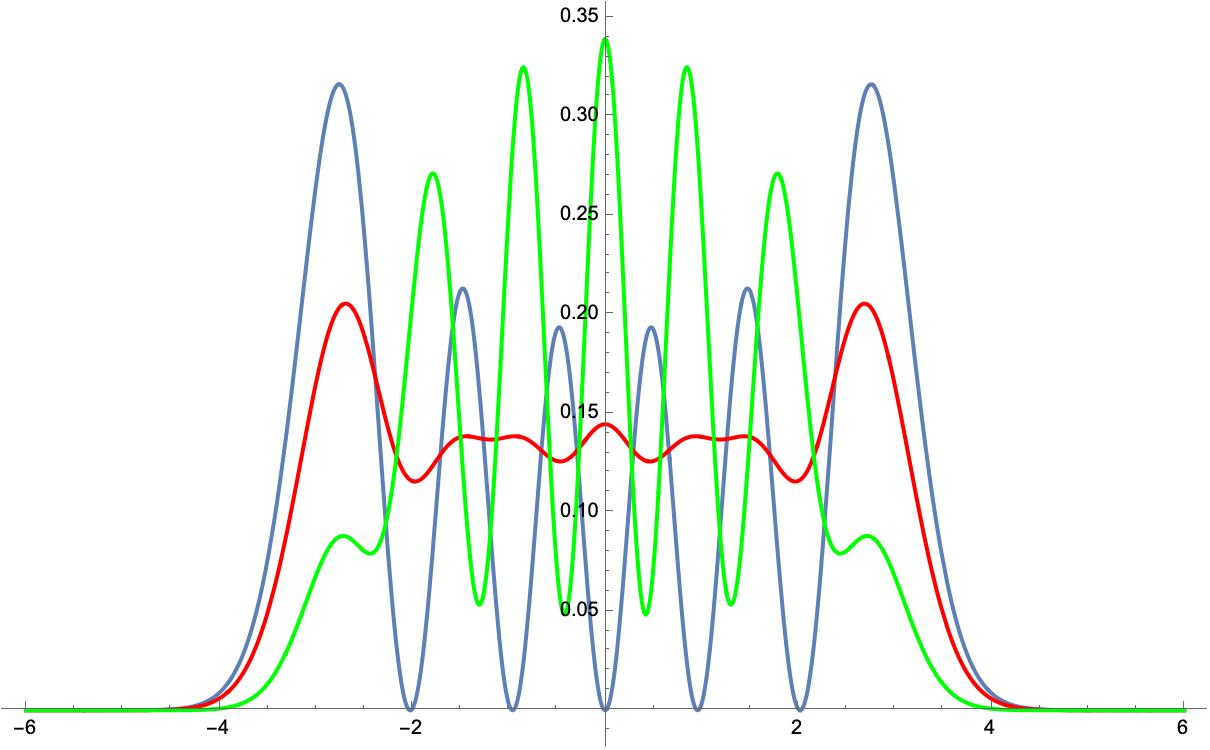} \\[\abovecaptionskip]
    \small (b) Fifth excited state
  \end{tabular}

  \caption{Corrected probability density of the harmonic oscillator from Eqn.\,(\ref{presence_density}), where the states are modified using the effective mass from Eqn.\,(\ref{eff_mass}). Here, $m = \omega = 1$, for $a = 0$ (blue), $a = 0.1$ (red), $a = 0.5$ (green).}
    
  \label{harmonic_osc}
\end{figure}

The corrected probability densities for the ground state and fifth excited state of the harmonic oscillator are displayed in Figure (\ref{harmonic_osc}).

Three qualitative features of this result, assuming $a>0$, deserve comment:
\begin{itemize}
\item The energy levels are no longer equally spaced.
\item The modifications of the energy levels become more significant as $n$ increases.
\item For $n \rightarrow \infty$, $\epsilon_n \rightarrow \infty$ and we have the leading behavior
\begin{equation}
E_n \rightarrow \frac{1}{2am}.
\end{equation}
The eigenvalues increase monotonically with $n$, but the discrete spectrum is bounded from above.
\end{itemize}

When the limiting energy for the discrete spectrum is exceeded, we have a continuous spectrum characterized by negative effective mass.  In this regime wave functions can respond to increases in potential by incorporating rapid oscillation.  Here we will present a heuristic discussion of that phenomenon.  For $m_{\rm eff.} > 0$  in the classically forbidden region $V > E$  the solution of 
\begin{equation}
\biggl(- \frac{1}{2m_{\rm eff}} \frac{d^2}{dx^2}  + V - E \biggr)  \psi  (x) ~=~ 0
\end{equation}
involves real exponentials $\psi (x ) \sim e^{\pm \sqrt{ 2m_{\rm eff.}  (V-E) } \, x }$.  Here the growing exponential behavior cannot easily be sustained in a normalizable  wave function, and thus we generally expect damping of amplitude in classically forbidden regions.  But if $m_{\rm eff.} < 0 $ the solutions are $\psi (x) \sim e^{\pm i \sqrt {2 |m_{\rm eff.}| \, (V-E)}\, x}$.  Thus they oscillate in the forbidden regions, and they oscillate more rapidly, the more forbidden those regions are.   This corresponds  to the classical behavior of a negative mass particle accelerating into a rising potential.  Related behavior is seen for positive mass in the inverted harmonic oscillator \cite{iho}.  It also has overtones of the Klein paradox, which we discuss briefly in Appendix \ref{Klein_paradox}.

Note that in the generalized Schr\"odinger equation it is $m_{\rm eff.}^{-1}$ that occurs naturally, and in this sense the difference between an infinitely positive and an infinitely negative effective mass is small.  Nevertheless it entails important qualitative consequences, as we shall see.  

\begin{figure}
  \centering
  \begin{tabular}{@{}c@{}}
    \includegraphics[width=10cm]{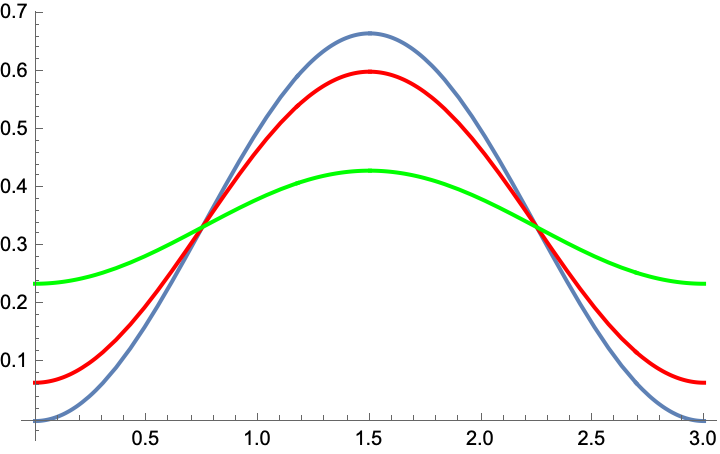} \\[\abovecaptionskip]
    \small (a) Ground state
  \end{tabular}

  \vspace{\floatsep}

  \begin{tabular}{@{}c@{}}
    \includegraphics[width=10cm]{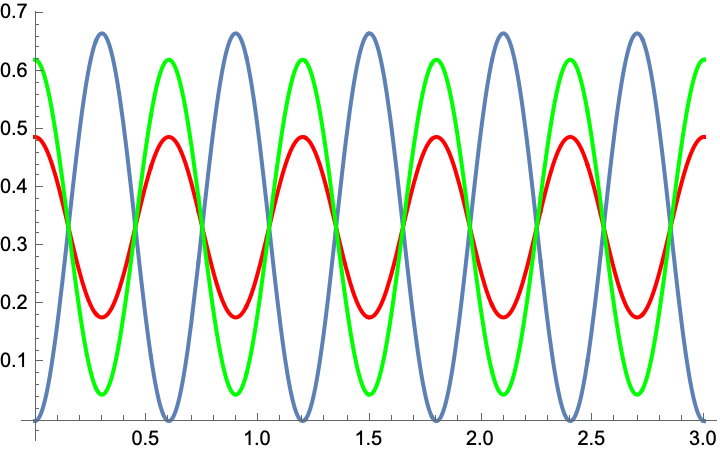} \\[\abovecaptionskip]
    \small (b) Fifth excited state
  \end{tabular}

  \caption{Corrected probability density of the infinite well, from Eqn.\,(\ref{presence_density}) with $L = 3$, for $a = 0$ (blue), $a = 0.1$ (red), $a = 0.5$ (green).}
  
  \label{infinite_well}
\end{figure}

For a particle confined within an infinitely deep well with length $L$, the energy eigenvalues of the conventional Schr\"odinger equation are
\begin{equation}
\epsilon_n ~=~ \frac{n^2 \pi^2}{2mL^2}
\end{equation}
with $n \geq 1$, and we infer for the modified Schr\"odinger equation the energy eigenvalues
\begin{equation}
E_n ~=~ \frac{n^2 \pi^2}{2mL^2}(1-2amE_n)
\end{equation}
leading to
\begin{equation}
E_n ~=~ \frac{\epsilon_n}{1+ 2am \epsilon_n}.
\end{equation}
The corrected probability densities for the ground state and fifth excited state of the infinite well are displayed in Figure (\ref{infinite_well}). Here too we find the limiting energy $\frac{1}{2am}$.  

The modified free-particle dispersion relation
\begin{equation}
\omega (1 + ak^2) ~=~ \frac{k^2}{2m}
\end{equation}
leads to the phase velocity
\begin{equation}
\frac{\omega}{k} ~=~ \frac{k}{2m} \frac{1}{1+ak^2} 
\end{equation}
and group velocity
\begin{equation}
\frac{d\omega}{dk} ~=~ \frac{k}{m} \frac{1}{(1 + ak^2)^2}.
\end{equation}
These indicate, for $a > 0$, a slowing of propagation. Indeed, we have a limiting phase velocity $\frac{1}{4m\sqrt a}$ and a limiting group velocity $\frac{3\sqrt 3}{16m\sqrt a}$. Both the wave and the group velocity approach 0 as $k \rightarrow \infty$.

Note here that the momentum associated to a plane wave $e^{i(kx -\omega t)}$  is $k(1+ a k^2)$, while its energy is $\omega$ as usual.

\subsection{Second Modified Lagrangian}\label{second_Lagrangian}

Now let us consider a second modified Lagrangian
\begin{equation}\label{second_modified_L}
L =  \frac{i}{2} \psi^* \stackrel{\leftrightarrow}D_t \psi + \frac{ia}{2}  \nabla \psi^*  \stackrel{\leftrightarrow}D_t \nabla \psi  - \frac{1}{2m} \nabla \psi^* \cdot \nabla \psi,
\end{equation}
where $D_t \equiv \partial_t - i A_0$ is the gauge covariant derivative.  This differs from Eqn.\,(\ref{modified_L}) in that the potential $A_0$ (corresponding to $V$) appears in the term proportional to $a$, as part of the covariant derivative.  Thus, Eqn.\,(\ref{second_modified_L}) supports a gauge symmetry 
\begin{eqnarray}
A_0 (x) ~&\rightarrow&~ A_0 (x) + \nabla f(x) \nonumber \\
\psi (x, t) ~&\rightarrow&~ e^{if(x)} \, \psi(x, t).
\end{eqnarray}
Thus, it is appropriate to consider in an effective theory of interactions of a charged particle (or field) represented by $\psi$ with the electric field $E = -\nabla A_0$.  

(Our first modified Lagrangian might represent the interactions of a charged particle (or field) represented by $\psi$ with other, gauge-invariant fields, such as phonon fields.  For electrons, whose primary interaction is with the electromagnetic field, these secondary interactions can be traced, ultimately, to electromagnetic interactions, but with fields or their gradients rather than potentials.  We focused on that case first, because the analysis of simple examples is, thanks to the emergence of an energy-dependent effective mass, more tractable.)

\subsubsection{Equations of Motion}

The modified Schr\"odinger equation is given by
\begin{align}
        0 &= \frac{\delta L}{\delta \psi^*} - \partial_t \frac{\delta L}{\delta \partial_t \psi^*} - \nabla \frac{\delta L}{\delta \nabla \psi^*} + \partial_t \nabla \frac{\delta L}{\delta \partial_t \nabla \psi^*}\notag\\
        &= -V\psi + \frac{i}{2} \partial_t \psi - \partial_t \biggl(-\frac{i}{2} \psi\biggr)\notag\\
        &- \nabla \biggl(\frac{ai}{2} \partial_t \nabla \psi - \frac{1}{2m} \nabla \psi - aV \nabla \psi\biggr) + \partial_t \nabla\biggl(-\frac{ai}{2} \nabla \psi\biggr)\notag\\
        &= -V \psi + i \partial_t \psi - ai \partial_t \nabla^2 \psi + \frac{1}{2m} \nabla^2 \psi + a\nabla(V \nabla \psi).
    \end{align}

\subsubsection{Conservation Laws}

\begin{enumerate}

\item We find the equation expressing local charge conservation
\begin{align}
        0 &= \partial_t \biggl(\psi^* \psi + a(\nabla \psi^* \cdot \nabla \psi)\biggr)\notag\\
        &+ \nabla\biggl(\frac{i}{2m}(\psi \nabla \psi^* - \psi^* \nabla \psi) - a(\psi \partial_t \nabla \psi^* + \psi^* \partial_t \nabla \psi)\notag\\
        &+ ai (\psi V \nabla \psi^* - \psi^* V \nabla \psi)\biggr).
        \end{align}
        
\item We can obtain an equation that expresses local energy conservation:
\begin{align}
        0 &= \partial_t \biggl(\psi^* V \psi + \biggl(aV + \frac{1}{2m}\biggr)(\nabla \psi^* \cdot \nabla \psi)\biggr)\notag\\
        &+ \nabla\biggl(ai(\partial_t \psi^* \partial_t \nabla \psi - \partial_t \nabla \psi^* \partial_t \psi) - \frac{1}{2m}(\partial_t \psi^* \nabla \psi + \nabla \psi^* \partial_t \psi)\notag\\
        &- a(V \nabla \psi \partial_t \psi^* + V \nabla \psi^* \partial_t \psi)\biggr)
    \end{align}
The formal expression of the energy density is given by
\begin{equation}
    \epsilon = \psi^* V \psi + \biggl(aV + \frac{1}{2m}\biggr)(\nabla \psi^* \cdot \nabla \psi)
    \end{equation}
which is now dependent on $a$, and the formal expression for its flux is
\begin{align}
       j_\epsilon &= ai(\partial_t \psi^* \partial_t \nabla \psi - \partial_t \nabla \psi^* \partial_t \psi) - \frac{1}{2m}(\partial_t \psi^* \nabla \psi + \nabla \psi^* \partial_t \psi)\notag\\
       &- a(V \nabla \psi \partial_t \psi^* + V \nabla \psi^* \partial_t \psi).
    \end{align}
    
\item We can obtain an equation that expresses the local change of momentum reacting to the force field $-\partial_k V$:
\begin{align}
        &-2(\psi \psi^* + a \nabla \psi \cdot \nabla \psi^*) \partial_k V \notag\\
        &= i\partial_t\biggl(\psi \partial_k \psi^* - \psi^* \partial_k \psi - a(\nabla \psi^* \partial_k \nabla \psi - \nabla \psi \partial_k \nabla \psi^*)\biggr)\notag\\
        &- \frac{1}{2m} \nabla(\psi^* \partial_k \nabla \psi - \nabla \psi \partial_k \psi^* + \psi \partial_k \nabla \psi^* - \nabla \psi^* \partial_k \psi)\notag\\
        &- ai \nabla(\partial_k \psi^* \partial_t \nabla \psi - \partial_k \psi \partial_t \nabla \psi^* + \psi \partial_k \partial_t \nabla \psi^* - \psi^* \partial_k \partial_t \nabla \psi)\notag\\
        &- a \nabla\biggl(\psi^* \partial_k(V \nabla \psi) - (V \nabla \psi) \partial_k \psi^* + \psi \partial_k(V \nabla \psi^*) - (V \nabla \psi^*) \partial_k \psi\biggr).
    \end{align}
Then, we can identify the momentum density
\begin{equation}
\pi_k = i\biggl(\psi \partial_k \psi^* - \psi^* \partial_k \psi - a(\nabla \psi^* \partial_k \nabla \psi - \nabla \psi \partial_k \nabla \psi^*)\biggr)
\end{equation}
and its flux
\begin{align}
j_l^{(\pi_k)} &= \frac{1}{2m} (\psi^* \partial_k \partial_l \psi - \partial_l \psi \partial_k \psi^* + \psi \partial_k \partial_l \psi^* - \partial_l \psi^* \partial_k \psi)\notag\\
&+ ai (\partial_k \psi^* \partial_t \partial_l \psi - \partial_k \psi \partial_t \partial_l \psi^* + \psi \partial_k \partial_t \partial_l \psi^* - \psi^* \partial_k \partial_t \partial_l \psi)\notag\\
&+ a \biggl(\psi^* \partial_k(V \partial_l \psi) - (V \partial_l \psi) \partial_k \psi^* + \psi \partial_k(V \partial_l \psi^*) - (V \partial_l \psi^*) \partial_k \psi\biggr).
\end{align}

\end{enumerate}

\subsubsection{Examples: Localization by Energy}

In a region where the potential $V$ is constant, stationary wave functions (i.e., wave functions with time dependence $\propto e^{-iEt}$) with wave vector $k$ (i.e., $\propto e^{ikx}$) satisfy
\begin{equation}\label{local_energy}
E-V ~=~ \frac{1}{2ma}\biggl( 1 - \frac{1}{1+ak^2}\biggr).
\end{equation}
Thus, propagating waves, with real $k$ and $k^2 \geq 0 $, occur only in the pass band
\begin{equation}
0 \leq E-V < \frac{1}{2ma}.
\end{equation}
This tight relationship between allowed energies and the local potential, which includes both upper and lower bounds, is unusual and has striking consequences.

If we have constant potentials $V_+$ for $x\geq 0$ and $V_-$ for $x<0$, with $| V_+ - V_- | > \frac{1}{2ma}$, then the world in effect divides into two separate parts, since the pass bands do not overlap.  Particles described by wave packets of any spatial form, incident from the left side, will be totally reflected, and likewise for particles incident from the right.  Note however that although the integrated probability for the reflected waves is unity, their form is not preserved.  

More generally, if we have piecewise constant potentials $V_j$ on the intervals $(x_j, x_{j+1})$ with no overlap among the local pass bands, i.e. $|V_j - V_k| > \frac{1}{2ma}$ for $j\neq k$, then we will have energy eigenfunctions for energy $E$ predominantly supported in the region with $|E - V_j | \leq \frac{1}{2ma}$; and particles, and a similar segregation of propagating states into separate worlds.  

If the potential varies slowly in space, we can apply the WKB approximation 
 \begin{equation}
 \psi (x, t) \, \approx \, A(x)\, \exp \biggl(i\int^x k(u)  \, du\biggr).
 \end{equation}
 In the first approximation, ignoring all gradients, we have an equation 
 \begin{equation}
 0 ~=~ \epsilon (1+ ak^2) - \frac{1}{2m}k^2 
 \end{equation}
 of the same form as Eqn.\,(\ref{local_energy}); we can also express it as
 \begin{equation}
 k^2 ~=~\frac{2m\epsilon}{1 - 2ma\epsilon}.
 \end{equation}
 The second approximation takes into account first-order gradients, in the form
\begin{equation} 
 0 \, = \, (1+ aV)(\nabla k \, A + 2k \nabla A) + ak \nabla V \, A
\end{equation}
with $k$ derived from the first approximation.  The validity of these approximations require that the wavelength does not vary significantly within a wavelength, i.e. 
\begin{equation}
1 ~>>~ \biggl|\frac{\nabla k}{k^2} \biggr|~=~ \biggl| \frac{\nabla V}{m^{1/2}(2\epsilon)^{3/2}(1-2ma\epsilon)^{1/2}} \biggr|.
\end{equation}
Thus it will fail not only near turning points $\epsilon \sim 0$, where $k \sim 0$, but also when $\epsilon \sim \frac{1}{2ma}$, where $k \sim \infty$.  This poses an interesting mathematical challenge, i.e. how to continue approximate WKB solutions through such danger zones, but we will not pursue it further here.

The phenomenon of localization by energy, discussed here, calls to mind Anderson localization, but it is quite different.  There is a subtle relationship, however, that deserves further investigation.  Specifically, for example, let us consider a periodic potential.  If the potential localization has large amplitude and long wavelength, the local pass bands corresponding to a definite energy will form a periodic array in space.  The wave function will be concentrated on those allowed regions, with exponential tails in the intervening forbidden zones.  These are the conditions under which a tight-binding model is an appropriate approximation, with small hopping terms.  We will have periodic, spatially extended solutions and a very flat band.  This is the usual {\it starting point\/} for an analysis of Anderson localization, which concerns the change in the spectrum and nature of the states in response to a small random potential.

\section{Hamiltonian and Local Energy}

To set up the Hamiltonian formalism in the most straightforward way, we take $\psi$ as the dynamical variable in Eqn.\,(\ref{modified_L}).  The terms linear in $\partial_t$, including the term proportional to $a$, do not contribute numerically to the Hamiltonian.  Note that our expression Eqn.\,(\ref{energy_density}) for energy density reflects this fact.  In particular, it is bounded below by the minimum of $V$.  Those terms do, however, control the identification of canonical momentum.  Indeed, we have
\begin{equation}
\pi_\psi ~\equiv \frac{\partial L}{\partial \partial_t \psi} ~=~ i (1 - a \nabla^2 ) \psi^*.
\end{equation}
(Here our comment at the end of Section \ref{point_structure} is relevant.)
To express the Hamiltonian in terms of $\pi_\psi$ we must invert this equation, in the form
\begin{equation}
\psi^* ~=~ -i (1-a \nabla^2)^{-1} \pi_\psi.
\end{equation}
This brings in non-locality, at a formal level.  Note that for $a \geq 0 $ the operator $1 - a \nabla^2$ is invertible, and can be expressed simply as a convolution in real space.

In the preceding discussion of examples we identified the $E$ that occurs in the factor $e^{-iEt}$ that accompanies stationary states with energy.  That identification is associated with the canonical pairing of energy and time, as realized in the abstract Schr\"odinger equation
\begin{equation}\label{abstract_schrodinger}
i \partial_t \psi ~=~ H \psi.
\end{equation}
Comparing Eqn.\,(\ref{modified_Schr}) with Eqn.\,(\ref{abstract_schrodinger}), we see that our modified Schr\"odinger equation fits into the abstract paradigm with the Hamiltonian operator
\begin{equation}
H ~=~ \frac{1}{1-a \nabla^2} (   V -  \frac{1}{2m} \nabla^2  ) .
\end{equation}

The issue arises, how to reconcile this expression for the Hamiltonian with the local energy density appearing in Eqn.\,(\ref{energy_density}).   The point is that in evaluating the energy density we must recognize that in view of the modified presence density Eqn.\,(\ref{presence_density}) the dual vector (``bra'') connected to the Hilbert space vector (``ket'') $\psi$ is no longer $\psi^*$, but rather
\begin{equation}
\langle \psi | ~=~ (1 - a \nabla^2) \psi^*.
\end{equation}
Thus, in the expectation value for energy density 
\begin{equation}
\langle \psi | H | \psi \rangle ~=~ \psi^* (V - \frac{1}{2m} \nabla^2) \psi,
\end{equation}
we recover Eqn.\,(\ref{energy_density}).  

\section{Discussion}

\begin{enumerate}

\item The $a$-term in Eqn.\,(\ref{modified_L}) is polynomial and of low mass dimension, and is therefore, in the spirit of Landau-Ginzburg theory, a very natural term to incorporate into the description of quasi-particles or emergent fields.  Since it is quadratic in the field (or, alternatively, wave function)  it affects free propagation directly, and brings in qualitatively new features.  Another quadratic term of a similar sort, {\it viz}. 
\begin{equation}\label{spin_momentum_term}
\Delta L_0 ~=~ b \psi^* \stackrel{\leftrightarrow}{\partial_t} \sigma \cdot \stackrel{\leftrightarrow} \nabla \psi
\end{equation}
can arise for two-component spinor fields, and is of even lower dimension.  It is rotationally invariant, but violates parity.  The candidate conserved ``probability of presence'' current associated to Eqn.\,(\ref{spin_momentum_term}) as it stands does not define a positive-definite inner product, so it is not suitable to provide a Hilbert space metric.  There are related terms, {\it e.g.}, 
\begin{equation}
\Delta L_1 ~=~ \psi^* i \stackrel{\leftrightarrow}{\partial_t} ( 1 + i b \sigma \cdot \stackrel{\leftrightarrow} \nabla)^2 \psi \, , 
\end{equation}
which are free of that difficulty.
\item  There is no difficulty in extending the preceding discussions to many-body wave functions and to models with more complex conventional interactions.   Thus we can infer modified equations of state for quantum ideal gases, modified densities of states, and so forth.  Second quantization brings in some formal novelties, but it is straightforward in principle.
\item Related to the preceding point: In the bulk of this paper we have used field Lagrangians to derive wave equations suitable to serve as models for individual quantum particles.  Our Lagrangians were quadratic in the field and supported a conserved quantum number, according to which the field carried a unit charge.  On the face of it, canonical quantization of field variables -- quantum field theory -- instructs us to treat field Lagrangians quite differently, generally leading to many-particle Hilbert spaces.  For the special class of Lagrangians under consideration, however, there is a consistent truncation to the one-particle sector, and it is governed by the equations we discussed.  
\item Terms of the kind we considered can also be added to Lagrangians of more general kinds, {\it e.g.} including terms of higher order in the field variables or not supporting a conserved quantum number, that must be treated using the full machinery of quantum field theory.  There is much room for further exploration in that direction.  Here let us observe that while many varieties of ``potential energy'' are considered in the literature of effective field theory, variant ``kinetic energy'' terms, involving time derivatives, have been comparatively neglected.
\item If we consider the path integral expression for transition amplitudes defined by $\int {\cal D} \psi^* {\cal D} \psi e^{iL + J(x, t) {\cal O} (x, t)}$, for various prescribed probes $J$ coupled to sources ${\cal O}$, we see that the expressions we obtained for densities, currents, and stresses have physical interpretations in line with their names.  This is essentially the logic of the Schwinger action principle \cite{schwinger}.  It gives local forms of the Hellman-Feynman theorem  \cite{deb} that can readily incorporate the possibility of the unconventional terms considered above, as we have shown directly.  
\end{enumerate}

\bigskip

\bigskip

{\it Acknowledgements\/}
We thank H. Hansson and Wu Biao for helpful comments.
FW is supported by the
U.S. Department of Energy under grant Contract 
Number DE-SC0012567 and by the Swedish Research 
Council under Contract No. 335-2014-7424. ZY is supported by a grant from the UROP office of MIT.  

\bigskip 

\bigskip

\appendix

\section{Non-relativistic Expansion of Dirac Equation}\label{Dirac_reduction}

Here we indicate how to obtain the non-relativistic expansion of the Dirac equation in Lagrangian form.  The Hamiltonian form is an immediate consequence.  Of course, this is an old subject, and there is nothing here that is essentially new, but we do want to demonstrate the soundness of our somewhat unorthodox perspective, as indicated in the first section of the text.   Also, it seems to us that existing presentations of the subject are more complicated than they need to be, so this brief, straightforward discussion may have pedagogical value. 

The Dirac equation for a unit charged particle, minimally coupled, reads
\begin{equation}
(i \gamma^\mu D_\mu - m) \, \psi ~=~ 0
\end{equation}
with the covariant derivative
\begin{equation}
D_\mu ~\equiv~ \partial_\mu - i A_\mu.
\end{equation}
To obtain a convenient non-relativistic limit, we use a representation of the $\gamma$ matrices in which $\gamma^0$ is as simple as possible and the spatial $\gamma^j$ respect its block structure, {\it viz}. 
\begin{eqnarray}
\gamma^0 ~&=&~ \left(\begin{array}{cc}1 & 0 \\0 & -1 \end{array}\right) \nonumber \\
\gamma^j ~&=&~ \left(\begin{array}{cc} 0 &  \sigma^j  \\ -\sigma^j & 0\end{array}\right).
\end{eqnarray}
Separating out the rapid time-dependence associated with rest mass energy, and writing the the 4-component spinor into top and bottom two-component spinors that live in the eigenspaces of $\gamma^0$, we define
\begin{equation}
\psi ~=~ e^{-imt} \, \left(\begin{array}{c}\phi \\ \eta \end{array}\right).
\end{equation}
In this notation, the Dirac equation reads
\begin{eqnarray}
i D_t \phi \, + \, i \vec{\sigma} \cdot \vec \nabla \, \eta ~&=&~0 \label{reduced_upper} \\
i \vec{\sigma} \cdot \vec \nabla \, \phi \, + \, (iD_t + 2m) \eta  ~&=&~ 0 \label{reduced_lower}.
\end{eqnarray} 

For a convenient non-relativistic limit we want equations that involve only two-component spinors, i.e. $\phi$.  To reach that goal, we can solve Eqn.\,(\ref{reduced_lower}) approximately to express $\eta$ in terms of $\phi$, and then insert the solution into Eqn.\,(\ref{reduced_upper}).  Appropriate approximations derive from an expansion in the parameter $\frac{D_t}{2m}$, according to
\begin{eqnarray}
\eta ~&=&~ (2m + i D_t) ^{-1} \, (- i \vec{\sigma} \cdot \vec \nabla \, \phi)  \nonumber \\
~&=&~ \biggl(\frac{1}{2m} \, - \, \frac{1}{4m^2} i D_t \, + \,  \frac{1}{8m^3} (i D_t)^2 \, + \, ... \biggr) (- i \vec{\sigma} \cdot \vec \nabla \, \phi)  \label{time_derivative_expansion}.
\end{eqnarray} 
The assumed smallness of $\frac{D_t}{2m}$ reflects that the wave function should not contain energies that differ too much from the rest mass (smallness of $\partial_t$) and also that the electric potentials $A_0$ should not become large compared to the rest mass.  Failure of the latter hypothesis brings in the physics of the Klein paradox, discussed in Appendix \ref{Klein_paradox}.

We get different approximations depending on how many terms we retain in Eqn.\,(\ref{time_derivative_expansion}).  Keeping only the first term, we get the Pauli equation
\begin{equation}\label{pauli_equation}
(i D_t \, + \, \frac{1}{2m} \nabla^2 \, +  \, \frac{1}{2m} \sigma \cdot B) \, \phi ~=~ 0 
\end{equation}
incorporating a spin magnetic moment with gyromagnetic ratio $g=2$.

The second order of approximation brings in 
\begin{equation}
- \frac{i}{4m^2} \,  \sigma \cdot \nabla  \, D_t  \, \sigma \cdot \nabla 
\end{equation}
acting on $\phi$.  If we set the vector potential to zero, this addition leads us to the equation we had in our ``second Lagrangian'' in the main text, with $a = \frac{1}{4m^2}$.  

Upon retaining the vector potential, we are faced with an elegant but rather inscrutable expression.
By re-arranging the pieces we can put it in a more transparent form.  Pulling $D_t$ to the left, we get 
\begin{equation}
- \frac{i}{4m^2} \,  \sigma \cdot \nabla  \, D_t  \, \sigma \cdot \nabla ~=~ - \frac{i}{4m^2} \,  D_t \, \sigma \cdot \nabla  \, \sigma \cdot \nabla + \frac{1}{4m^2} \, \sigma \cdot E \, \sigma \cdot \nabla
\end{equation}
where the second term on the right arises from the commutator of covariant derivatives.  The first term on the right then involves the same structure we saw in the first approximation, {\it viz}. 
\begin{equation}
- \frac{i}{4m^2} \,  D_t \, \sigma \cdot \nabla  \, \sigma \cdot \nabla ~=~ - \frac{i}{4m^2} \,  D_t \, (\nabla^2  \, + \, \sigma \cdot B)
\end{equation}
while the second term, upon similarly expanding the product of $\sigma$ matrices, becomes
\begin{equation}
 \frac{1}{4m^2} \, \sigma \cdot E \, \sigma \cdot \nabla ~=~  \frac{i}{4m^2} (-iE \, \cdot \nabla + \sigma \cdot (E \, \times \, \nabla).
\end{equation}
Altogether then, the second approximation brings in
\begin{equation}
\frac{i}{4m^2} \,  \bigl(- D_t \, (\nabla^2  \, + \, \sigma \cdot B) \, - \, i E \, \cdot \nabla + \sigma \cdot (E \, \times \, \nabla) \bigr).
\end{equation}

The third order of approximation brings in second order time derivatives, and so changes the initial value problem.  Thus, in order to evolve the $\phi$ into the future we must specify $\partial_t \phi$ as well as $\phi$ itself at the initial time.  We can do this, for example, by imposing the second-order approximation to define the initial time derivative.  We will not pursue that further here. It is straightforward in principle to put the equations into a Lagrangian and Hamiltonian framework, but complicated in practice.  

The fourth order of approximation brings in third order time derivatives.  This introduces new degrees of freedom, basically reinstating $\eta$ as an independent field.  At this point, if not before, the utility of the expansion has become dubious.  

To implement quantum theory it is important to have Lagrangian and Hamiltonian formulations of the equations of motion.  Above, up to the second order, we have equations linear in time derivatives, and so appropriate Lagrangians and Hamiltonians are immediately at hand.  Indeed, the schematic equation
\begin{equation}
f(x) \partial_t \phi + g(x) \phi ~=~ 0
\end{equation}
follows from the Lagrangian
\begin{equation}
L ~=~ \phi^\dagger (f \partial_t + g) \phi
\end{equation}
and the Hamiltonian numerically equal to
\begin{equation}
H ~=~ - \phi^\dagger g \phi.
\end{equation}
Note however that the canonical conjugate to $\phi$ is $ f \phi^\dagger$.  As a consequence the canonical commutation relations and the expression for the Hamiltonian in terms of canonical variables are unusual, as we have encountered and explained in the main text.

\section{Klein Paradox}\label{Klein_paradox}

The Klein paradox \cite{klein} dates from the earliest consideration of relativistic wave equations for quantum theory.  It played an important historical role in the development of quantum field theory.  In recent years it has attracted renewed research interest, especially in the context of graphene, Weyl semi-metals, and related materials.  There is a large literature on the subject; notable recent reviews include \cite{hansen, dombey, katsnelson}.  

Klein originally considered the application of the Dirac equation to an electron incident from the left on a potential step  
\begin{equation}
V(x) ~=~ H(x) V_0
\end{equation}
where $H$ is the Heaviside function.  He found the paradoxical result that for a strong repulsive potential, $eV_0 > 2m$, the reflection coefficient (calculated straightforwardly) is greater than unity, and the transmission coefficient is negative.  Klein, acknowledging input from Pauli, suggested a way to repair the problem.  By demanding that the current for $x>0$ flows to the right, one is led to choose a negative value for the momentum on that side.  With that choice, one finds positive values for both transmission and reflection, with the sum equal to unity.  An interesting feature of the result is that the transmission approaches a finite limit for $V_0 \rightarrow \infty$ with $E/V$ finite.   This surprising phenomenon, that there is significant propagation into the classically forbidden region, is known as ``Klein tunneling''.  

Since the Dirac equation supports a conserved current with a positive-definite probability density, we can maintain a one-particle interpretation consistently.  In this interpretation, however, we must allow both positive and negative energy solutions.  In the Klein set-up, one has the circumstance that positive energy $E >m $ on the left-hand side ($x<0$) can correspond to negative local energy $-(E-V_0) < -m $ on the right-hand side.  The (locally) negative-energy solution involves the lower components of the four-spinor, i.e. $\eta$ in the notation of Appendix \ref{Dirac_reduction}, so the reversal of current associated with a given $\vec p$, as advocated by Klein and Pauli, is appropriate.  

Nevertheless Klein's paradox points to a striking tension in the straightforward interpretation of the Dirac equation as a single-particle wave equation.   With the realization that the physically correct treatment of the Dirac equation brings in the existence of antiparticles, it became clear that a relevant physical effect was missing from Klein's analysis.   Namely, in the presence of the large step potential it is energetically favorable to produce an electron-positron pair: With the electron in the high-potential region and the positron in the low-potential region, the gain in potential energy is more than enough to compensate for the rest masses.  Thus, it is physically inconsistent to interpret the step potential together with the no-particle state as a static background, upon which we study propagation of electrons.  

We can come realize a stationary situation that corresponds to the Klein problem by turning on the potential adiabatically.  We must also maintain it by external means, since the produced pairs will tend to discharge it.   If we set up the problem that way, then the time-independent state containing the potential will also contain a steady flow of electrons to the right and positrons to the right, as the potential emits sparks.  (Note that this state does not respect time-reversal symmetry; i.e., it is stationary but not static.)    We must then take into account the effect of this background on the incident electron.  This is worked out nicely, taking into account the effect of quantum statistics, in Hansen and Ravndal \cite{hansen}.   There are also effects from proper electromagnetic interactions, beyond quantum statistics.  These effects can become very important and change the physics qualitatively, as we will discuss elsewhere \cite{prep}. 

The foregoing remarks are pertinent  more generally, in problems that involve electrons (or quasi-particles) interacting with strong external potentials.  

As emphasized by Dombey and Calogeracos \cite{dombey}, the phenomena of Klein tunneling and pair production are logically distinct.   Notably, for example,  ``paradoxical'' enhanced tunneling also occurs for sub-critical fields ($V_0 < 2m$).  In general, in modeling the behavior of electrons (or quasi-particles), we must be sure that we specify the physical situation we are modeling adequately.  Thus, in considering the behavior of electrons (or quasi-particles) in the presence of a large fixed electromagnetic potential, we must take into account the necessity of maintaining the potential, the possibility of continuing pair creation, and the statistical and electromagnetic influence of the produced pairs.   Having said that, the one-particle interpretation of the Dirac equation is often a valid first approximation, and Klein tunneling is a legitimate physical consequence of it.  In the words of Dombey and Calogeracos, 
\begin{quote}
Klein tunneling is a property of relativistic wave equations and not necessarily connected to particle emission.
\end{quote} 

Taking that insight a step further, in the body of this paper we have demonstrated that ``paradoxical'' behavior at least loosely related to Klein tunneling, notably including energy-dependent effective masses that can become negative and the phenomenon of localization by energy, can occur even in non-relativistic equations inspired by the Dirac equation, but allowing more flexible parameters.   For these equations, as we have emphasized, the one-particle interpretation is fully consistent with the general principles of quantum theory, and the question of antiparticle production does not arise.

\end{document}